\documentclass[a4paper,12pt]{article}
\usepackage[utf8]{inputenc}

\usepackage{amsmath,amsfonts,amsthm,amssymb,dsfont}
\usepackage{graphicx,wrapfig,lipsum}
\usepackage[section]{placeins}
\usepackage{indentfirst}

\usepackage[numbers,sort&compress]{natbib}

\usepackage{tikz}
\usetikzlibrary{shapes.geometric,decorations.markings}
\usepackage{subfigure}

\usepackage[hidelinks]{hyperref}

\hypersetup{
    colorlinks=true,
    linktocpage=true,
    linkcolor=blue,
    urlcolor=blue,
    citecolor=blue,
}

\numberwithin{equation}{section}
\newcommand{\Nc}{N_{c}}
\newcommand{\tNc}{\tilde{N}_{c}}
\newcommand{\Nf}{N_{f}}
\newcommand{\bQ}{\bar{Q}}
\newcommand{\Tr}{\text{Tr}}
\newcommand{\vev}[1]{\langle #1 \rangle}
\newcommand{\None}{{\cal N}=1}
\newcommand{\N}{\mathcal{N}}
\newcommand{\Mij}{M^{i}_{\phantom{i}j}}
\newcommand{\calD}{\mathcal{D}}

\def\beq{\begin{equation}}
\def\eeq{\end{equation}}
\newcommand{\bea}{\begin{eqnarray}}
\newcommand{\eea}{\end{eqnarray}}
\def\bal{\begin{align}}
\def\eal{\end{align}}

\begin{document}

\begin{titlepage}

\begin{center}

$\phantom{.}$\\ \vspace{2cm}
\noindent{\Large{\textbf{A Note on Seiberg Duality for $N_f < N_c$}}}

\vspace{1cm}

Adi Armoni \footnote{a.armoni@swansea.ac.uk} and Ricardo Stuardo \footnote{ricardostuardotroncoso@gmail.com}

\vspace{0.5cm}

\textit{Department of Physics, Faculty of Science and Engineering\\
        Swansea University, SA2 8PP, UK}

\end{center}

\vspace{0.5cm}
\centerline{\textbf{Abstract}} 

\vspace{0.5cm}

\noindent{We consider SQCD for $N_f<N_c$. We use a brane configuration in type IIA string theory to propose a Seiberg dual. The brane configuration that realises the magnetic theory contains $N_f$ branes and $N_c-N_f$ \textit{anti}-branes, leading to an $SU(N_c-N_f)$ gauge theory with $N_f$ flavours. The potential between the branes triggers a RG flow and we argue that the IR theory realises the Affleck-Dine-Seiberg superpotential. We use both field theory arguments, in particular anomaly matching, and a lift of the type IIA brane configuration to M-theory, in order to support our proposal.}

\vspace*{\fill}

\end{titlepage}

\newpage

\tableofcontents
\thispagestyle{empty}

\newpage
\setcounter{page}{1}
\setcounter{footnote}{0}

\section{Introduction}

Supersymmetry is the most powerful analytic tool to study the strong coupling regime of gauge theories, due to the contrainsts imposed by holomorphicity \cite{Intriligator:1995au}. $\None$ supersymmetry in 4d is sufficient to determine certain condensates that tell us about the vacua of the theory.

One of the most fascinating results in supersymmetric 4d gauge theories is Seiberg duality \cite{Seiberg:1994pq}: two distinct SQCD theories flow in the IR to the same fixed point. The duality had been extended to lower dimensions \cite{Giveon:2008zn}, various gauge groups and even to the case without supersymmetry \cite{Armoni:2013ika}.

The original version of the duality relates an 'electric' $SU(N_c)$ SQCD with $N_f>N_c$ to a 'magnetic' theory based on $SU(N_f-N_c)$ group. The case with $N_f<N_c$ does not seem to admit a sensible dual, since the theory with massless quarks admits a runaway Affleck-Dine-Seiberg (ADS) superpotential, namely no supersymmetric vacuum. In this paper we address the question whether if it is possible to make sense of a duality of SQCD with $\Nf < \Nc$,  namely if non perturbative results in the electric theory can be recovered using a magnetic dual.

In previous studies, in a 3d setup, when the dynamics is simpler, we argued that the answer is positive \cite{Armoni:2023ohe}. Using a brane setup we engineered a magnetic theory and demonstrated that both the electric and the magnetic dual admit the same vacuum even when $N_f +|k| <N_c$\footnote{In 3d Giveon-Kutasov duality \cite{Giveon:2008zn} relates $U(N_c)_k$ Yang-Mills Chern-Simons theory to a similar theory with a gauge group $U(N_f+|k|-N_c)_k$, with $k$ the CS level.}. In this paper we propose a magnetic dual of 4d SQCD with $N_f<N_c$. We use a brane setup to find the magnetic theory. The brane setup, as in 3d, contains both brane and \textit{anti}-branes, resulting in the ADS superpotential. The idea that a brane configuration with branes and \textit{anti}-branes realises dynamical supersymmetry breaking is not new: it was used in various papers in the past, in particular by Maldacena and Nastase \cite{Maldacena:2001pb} and in \cite{Garcia-Etxebarria:2012ypj,Armoni:2023ohe}, in both 3d and 4d. 

Let us elaborate on the basic idea: in ref.\cite{Elitzur:1997fh} Elitzur, Giveon and Kutasov proposed that Seiberg duality is relaised by a fivebrane swap in type IIA string theory. Their proposal passed many tests and admitted numerous generalisations. Let us swap the fivebranes when $N_f<N_c$: the resulting magnetic brane configuration contains $N_c-N_f$ \textit{anti} D4 branes, leading to $SU(N_c-N_f)$ gauge theory. Due to the absensce of supersymmetry we expect a potential between the colour branes and the flavour branes. 
The long distance potential is dictated by massless closed strings. In a brane anti-brane system both NS-NS and R-R closed string exchange lead to attraction. The short distance potential is dictated by massless open strings, namely field theory. The ADS superpotential is repulsive.
As a result the branes will settle at a minimum of the potential. While it is difficult to establish the precise location of the minimum we will use field theory to argue that the distance between the flavour and colour branes, where the branes settle, is inverse proportional to the electric quark mass.
We demonstrate our claim by lifting the electric theory to M-theory and carrying out the duality transformation. 

Our proposal passes several tests: first, the same idea was successfully used in 3d. We may consider the 3d theory as the dimensional reduction of the 4d theory. Moreover, we show that that the dual pair admits the same global symmetries and 't Hooft anomalies. Finally, we show that the M-theory lift of the magnetic theory describes the ADS superpotential.

The paper is organised as follows: in order to motivate our proposal we review in section 2 results from 3d. In section 3 we review SQCD dyanmics for $N_f < N_c$. In section 4 we use branes to propose a Seiberg duality for $N_f<N_c$. Section 5 is devoted to field theory checks of the duality. In section 6 we provide M-theory of the duality. In section 7 we discuss the meaning of the duality and propose future directions of research.

\section{Review of the 3d QCD Dualities From Branes}

Let us start by reviewing the results of \cite{Armoni:2022wef}, which recovers the three phases of adjoint QCD in 3d \cite{Gomis:2017ixy} by using brane dynamics. 

The starting point is to consider brane configurations of D3-D5-NS5 branes which have a low energy worldvolume 3d $\mathcal{N}=2$ $U(\Nc)_{k',k'}$ SYM. We refer to this theory as the electric theory. A similar brane configuration realises the magnetic dual. Furthermore, we are interested in the case\footnote{It is also possible to start with Chern-Simons level $k''$ with $k''<0$. We refer to this theory as electric' and its magnetic dual as magnetic'. While the electric and electric' theories are the same the resulting magnetic and magnetic' cover different patches of the phase diagram.} $k'>0$ and $k'<\Nc$. The directions in which the branes are extended are depicted in table \ref{3DN=2}. 

\begin{table}[h]
    \centering
    \begin{tabular}{c|c|c|c|c|c|c|c|c|c|c|}
           &  $x^{0}$ & $x^{1}$ & $x^{2}$ & $x^{3}$ & $x^{4}$ & $x^{5}$ & $x^{6}$ & $x^{7}$ & $x^{8}$ & $x^{9}$   \\ \hline
       D3  & - & - & - & $\cdot$ & $\cdot$ & $\cdot$ & - & $\cdot$ & $\cdot$ & $\cdot$  \\
       NS5 &  - & - & - & - & - & - & $\cdot$ & $\cdot$ & $\cdot$ & $\cdot$ \\
       NS5' &  - & - & - & - & $\cdot$ & $\cdot$ & $\cdot$ & $\cdot$ & - & - \\
       D5 &  - & - & - & $\cdot$ & $\cdot$ & $\cdot$ & $\cdot$ & - & - & -
    \end{tabular}
    \caption{Brane Configuration for the 3d $\N=2$ SYM theory.}
    \label{3DN=2}
\end{table}   

For the electric theory, we consider $\Nc$ D3-branes suspended between a NS5-brane on the left, and a NS5'-brane on the right, with $k'$ D5-branes in between the two Neveu-Schwarz branes. We can bring the $k'$ D5-branes close to the NS5'-brane to create a (1,$k'$) tilted fivebrane, realising the Chern-Simons term in the brane worldvolume. 

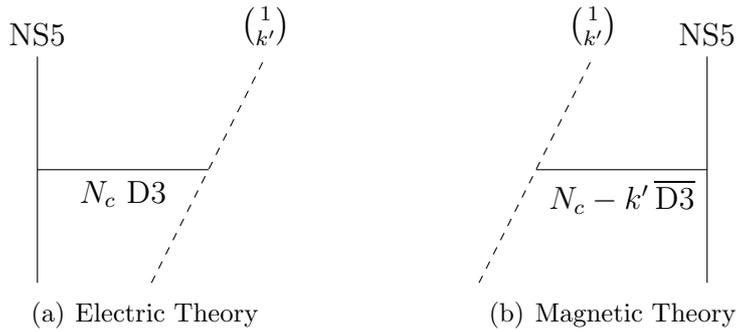
\begin{figure}[h!]
    \centering
    
    %
    \subfigure[Electric Theory]{
    \begin{tikzpicture}
    
    \draw (-1.5,-1.5) -- (-1.5,1.5) node [above] {NS5};

    \draw (-1.5,0) -- (0.75,0) node [midway,below] {$N_{c}$ D3};

    \draw[dashed] (0,-1.5) -- (1.5,1.5) node [above] {$\binom{1}{k'}$};   
    \end{tikzpicture}}
    \quad\quad\quad\quad\quad%
    %
    %
    \subfigure[Magnetic Theory]{
    \begin{tikzpicture}
    \draw (1.5,-1.5) -- (1.5,1.5) node [above] {NS5};

    \draw (-0.75,0) -- (1.5,0) node [midway,below] {$\Nc-k' \, \overline{\text{D3}}$};

    \draw[dashed] (-1.5,-1.5) -- (0,1.5) node [above] {$\binom{1}{k'}$};
    
	\end{tikzpicture}
    }
    \caption{The electric and magnetic 3d $\N=2$ brane configurations.}
        \label{BraneConfigurations3D}
 \end{figure}

To obtain the magnetic theory, we follow the prescription of \cite{Giveon:2008zn}. We obtain $(k'-\Nc)$ D3-branes suspended between the tilted fivebrane on the left and the NS5-brane on the right. Since we are considering $k'<\Nc$, it seems that there is a negative number of D3-branes on the magnetic configuration. In \cite{Armoni:2022wef}, this is interpreted as $(\Nc-k')$ \textit{anti}-D3-branes. Both brane configurations are depicted in figure \ref{BraneConfigurations3D}.

There is an interesting issue with the sign of the Chern-Simons term in the magnetic theory. Notice that since the tilted fivebrane is on the opposite side with respect to the electric theory, the sign of the Chern-Simons level changes. However, since we are considering \textit{anti}-branes the sign of the level changes again. Thus we obtain $U(\Nc-k')_{k',k'}$ SYM as the magnetic theory. 

\subsection{Phases of adjoint \texorpdfstring{QCD$_{3}$}{QCD3} from duality}

To obtain adjoint QCD$_{3}$, we notice that the $\N=2$ vector multiplet $V=(A_{\mu},\lambda_{1},\lambda_{2},\phi)$ can be written in terms of $\N=1$ variables as vector multiplet $\mathcal{V}=(A_{\mu},\lambda_1)$ and a scalar multiplet $\Phi=(\lambda_{2},\phi)$. Therefore, by integrating out the fermion in the scalar multiplet, we obtain the desired theory.

Let us start with the electric theory. We choose $k' = k + \frac{\Nc}{2}$. The conditions $k'>0$ and $k'<\Nc$ lead to $|k|<\frac{\Nc}{2}$. Then, by giving a large negative mass to the fermion in the scalar multiplet and integrating it out leads to $SU(\Nc)_k$. On the other hand, had we started with the electric' theory, instead we choose $k''=k-\frac{\Nc}{2}$. As before, the conditions $k''<0$ and $|k|''>-\Nc$ lead to $|k| < \frac{\Nc}{2}$. Integrating out the adjoint fermion in the scalar multiplet for large positive mass results again in $SU(\Nc)_k$. 

On the magnetic theory, after giving a large positive mass to the fermion in the scalar multiplet and integrating it out we get 
\begin{equation}
        U(N_c/2-k)_{\frac{3N_c}{4} + \frac{k}{2}, N_c}.
    \end{equation}
Similarly for the magnetic' theory for large positive mass
\begin{equation}
        U(N_c/2+k)_{-\frac{3N_c}{4} + \frac{k}{2}, -N_c} \, .
    \end{equation}
Both the magnetic and the magnetic' theory were conjectured in \cite{Gomis:2017ixy}. A very interesting outcome of the proposed duality is the existence of an intermediate quantum phase where the IR dynamics of the electric theory is a $U(N_c/2-k)_{\frac{N_c}{2} +k, N_c}$ TQFT. 

The branes were useful to reveal the infra red dynamics of a strongly coupled gauge theory including a phase which is not described by a semi-classical dynamics.

The surprise is that the duality was invoked for a range of parameters analogous to $N_f<N_c$ in 4d. That fact is the main motivation behind this paper.

\section{Brief Review of 4d \texorpdfstring{$\mathcal{N}$}{N}=1 SQCD with \texorpdfstring{$\Nf<\Nc$}{Nf<Nc}}

Let us now turn to the main subject of study of this paper, that is $\mathcal{N}=1$ $SU(\Nc)$ SQCD with $\Nf$ flavours in the (anti-)fundamental representation. We start by reviewing some field theory results. 

The global symmetries and charges are given in table \ref{ChargesElectric2} below. 
\begin{table}[h]
        \centering
        \begin{tabular}{c|cccc}
                  &  $SU(\Nf)_{L}$ & $SU(\Nf)_{R}$ & $U(1)_{B}$ & $U(1)_{R}$  \\ \hline 
            $Q$   &     $\Nf$      &    $1$        &    $1$     & $\frac{\Nf-\Nc}{\Nf}$ \\
            $\bQ$ &      $1$       &  $\bar{N}_{f}$  &    $-1$    & $\frac{\Nf-\Nc}{\Nf}$ 
        \end{tabular}
        \caption{Matter content and charges of the Electric Theory}
        \label{ChargesElectric2}
    \end{table}

While at tree level there is no superpotential for $\Nf<\Nc$, a non-perturbative superpotential is dynamically generated \cite{Affleck:1983mk}. In terms of $\Mij = Q^{i}\bar{Q}_{j}$ the ADS superpotential is given by
    \begin{equation}
        W_{ADS} = (\Nc-\Nf)\left( \frac{\Lambda^{3\Nc-\Nf}}{\det(M)} \right)^{\frac{1}{\Nc-\Nf}}.
    \end{equation}
This superpotential is generated by gaugino condensation for $\Nf<\Nc-1$ and by instantons for $\Nf=\Nc-1$. The squark potential,
    \begin{equation}\label{potential}
        V_{Q\bar{Q}} \sim \sum^{\Nf}_{i =1} \left| \frac{\partial W}{\partial Q} \right|^{2} \sim |M|^{-\frac{2\Nc}{\Nc-\Nf}}, 
    \end{equation}
is runaway, that is, the minimum of the potential is located at $\langle M \rangle \rightarrow + \infty$. SUSY is hence spontaneously broken: there is a SUSY vacuum but it is never reached. 

In order to stabilise the vacuum we may add masses to the chiral multiplet by modifying the superpotential as
    \begin{equation}
        W_{eff} = W_{ADS} + m_{i}^{\phantom{i}j} \Mij.
    \end{equation}
Upon the inclusion of these mass terms, the superpotential admits a minimum at a finite value of $\langle \Mij \rangle$, which can be obtained by extremising the superpotential
    \begin{equation}\label{VacuumWithMass}
        \langle \Mij \rangle = \left( \det(m) \Lambda^{3\Nc-\Nf}\right)^{\frac{1}{\Nc}} \left( m^{-1} \right)^{i}_{\phantom{i}j}.
    \end{equation}

The same behaviour can be obtained by starting from $\mathcal{N}=2$ SQCD and giving a mass, $\mu$, to the adjoint scalar. This mass deformation breaks $\mathcal{N}=2 \rightarrow \mathcal{N}=1$, and by integrating out the massive adjoint scalar an effective superpotential for $\Mij$ is generated
    \begin{equation}
        \Delta W = \frac{1}{2\mu}\left( \Tr(M)^{2} - \frac{1}{\Nc}(\Tr(M))^{2} \right).
    \end{equation}

By extremizing $\bar{W}_{eff} = W_{eff} + \Delta W$ and taking the $\mu\rightarrow +\infty$ limit, one recovers \eqref{VacuumWithMass}\footnote{In fact, the duality we propose holds even when $\mu$ is finite.}.

From now on we work in a basis in which $m^{i}_{j} = \text{diag}(m_{1},...,m_{\Nf})$ and $\Mij = \text{diag}(M_{1},...,M_{\Nf})$. In this basis, \eqref{VacuumWithMass} can be written as
    \begin{equation}\label{vevM}
        \vev{M}_{i} = \frac{\left(\prod_{j}m_{j}\right)^{\frac{1}{\Nc}}\Lambda^{\frac{3\Nc-\Nf}{\Nc}}}{m_{i}} = \frac{\calD}{m_{i}},
    \end{equation}
where we defined
    \begin{equation}
        \calD = \left(\prod_{j}m_{j}\right)^{\frac{1}{\Nc}}\Lambda^{\frac{3\Nc-\Nf}{\Nc}}.
    \end{equation}
This definition will be used later in the paper.

\section{Duality from Brane Dynamics}

We would like to generalise Seiberg duality \cite{Seiberg:1994pq} to the case in which the magnetic theory is non-SUSY, that is, when the number of colours $\Nc$ is greater than the number of flavours $\Nf$ ($\Nf<\Nc$). To this end, we use type IIA String Theory. The brane configuration for the electric and magnetic theory is the same as in the SUSY case \cite{Elitzur:1997fh}: we consider the configuration of D4, NS5 and D6-branes as given in the following table
\begin{table}[h]
    \centering
    \begin{tabular}{c|c|c|c|c|c|c|c|c|c|c|}
           &  $x^{0}$ & $x^{1}$ & $x^{2}$ & $x^{3}$ & $x^{4}$ & $x^{5}$ & $x^{6}$ & $x^{7}$ & $x^{8}$ & $x^{9}$   \\ \hline
       D4   & - & - & - & - & $\cdot$ & $\cdot$ & - & $\cdot$ & $\cdot$ & $\cdot$  \\
       NS5  & - & - & - & - & - & - & $\cdot$ & $\cdot$ & $\cdot$ & $\cdot$ \\
       NS5' & - & - & - & - & $\cdot$ & $\cdot$ & $\cdot$ & $\cdot$ & - & -\\
       D6   & - & - & - & - & $\cdot$ & $\cdot$ & $\cdot$ & - & - & - 
    \end{tabular}
\end{table}

In the electric theory $\Nc$ D4 (colour) branes are suspended, in the $x^{6}$ direction, between the NS5-brane (on the left) and the NS5'-brane (on the right). Also, we have $\Nf$ D6-branes are located on the left of the NS5-brane with D4 (flavour) branes suspended between the D6-brane and the NS5-brane. The effective theory on the D4 colour branes is $\mathcal{N}=1$ $SU(\Nc)$ SYM with $\Nf$ flavours in the fundamental representation. For the case of interest $\Nf<\Nc$, although the brane configuration is SUSY, the effective theory for massless quarks does not have a SUSY vacuum due to the generation of the non-perturbative ADS superpotential.

To obtain the brane configuration of the magnetic theory, we follow  \cite{Elitzur:1997fh} and swap the positions of the NS5 and NS5' branes. The resulting configuration consists of $\Nf$ D6-branes with D4-branes suspended between the D6 and the NS5' brane, and $\Nf-\Nc$ (colour) branes suspended between the NS5'-brane (on the left) and the NS5-brane (on the right). Since in our case $\Nf-\Nc<0$, it might seem that the effective theory on the colour D4-branes is a YM theory with negative rank $SU(-(\Nc-\Nf))$. The main idea in this paper is that one should instead think of the $\Nf-\Nc$ D4-branes as $\tNc=\Nc-\Nf$ \textit{anti}-D4-branes ($\bar{\text{D4}}$) with gauge group $SU(\Nc-\Nf)$.

Note that in the magnetic theory, since the NS5' and the D6-branes share the $(x^{8},x^{9})$ directions, allowing excitations of the $\Nf$ D4-branes along those directions, realising the meson field in the magnetic theory. 

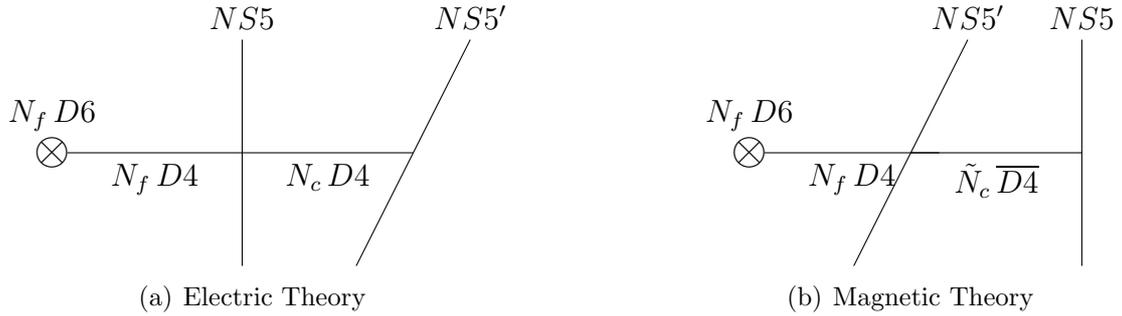
\begin{figure}[h!]
    \centering
    
    %
    \subfigure[Electric Theory]{
    \begin{tikzpicture}
    
    \draw (-1.5,-1.5) -- (-1.5,1.5) node [above] {$NS5$};

    \draw (-1.5,0) -- (0.75,0) node [midway,below] {$N_{c} \, D4$};

    \draw (0,-1.5) -- (1.5,1.5) node [above] {$NS5'$};

    \draw (-3.8,0) -- (-1.5,0) node [midway,below] {$N_{f} \, D4$}; 
    
    \draw (-4,0.15) node[above] {$\Nf \, D6$};
    \node at (-4,0){\Large $\otimes$};
   
    \end{tikzpicture}}
    \quad\quad\quad\quad\quad%
    %
    %
    \subfigure[Magnetic Theory]{
    \begin{tikzpicture}
    \draw (1.5,-1.5) -- (1.5,1.5) node [above] {$NS5$};

    \draw (-0.75,0) -- (1.5,0) node [midway,below] {$\tilde{N}_{c} \, \overline{D4}$};

    \draw (-1.5,-1.5) -- (0,1.5) node [above] {$NS5'$};

    \draw (-2.675,0) -- (-0.375,0) node [midway,below] {$N_{f} \, D4$};
    
    \draw (-2.875,0.15) node[above] {$\Nf \, D6$};
    \node at (-2.875,0){\Large $\otimes$};
    
	\end{tikzpicture}
	\label{MagSetUp}
    }
    \caption{Dualities from Branes}
    \label{BraneConfiguration}
	\end{figure}

The brane configurations for the electric and magnetic theories are given in figure \ref{BraneConfiguration} below

The effective theory on the \textit{anti}-brane is also $\mathcal{N}=1$ $SU(\Nc-\Nf)$ SYM: the interaction with the flavour branes that breaks supersymmetry explicitly. This is clear from the brane configuration: we have branes and \textit{anti}-branes. We claim below that the interaction between them in the magnetic theory displaces the flavour branes to infinity, which is the magnetic counterpart of the ADS superpotential behaviour.

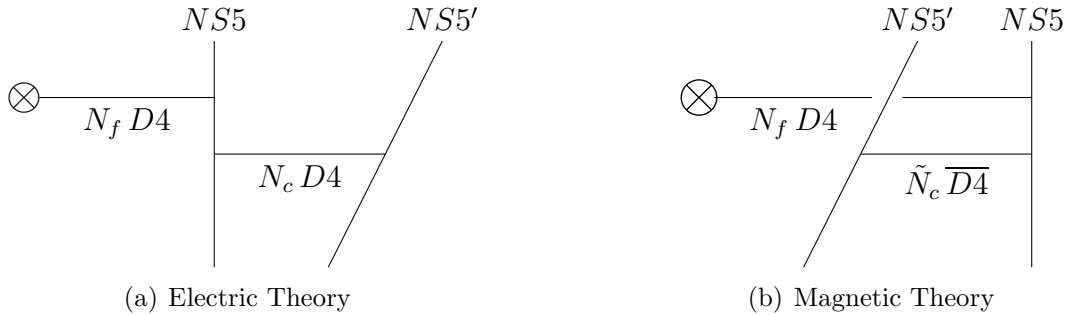
\begin{figure}[h!]
    \centering
    
    %
    \subfigure[Electric Theory]{
    \begin{tikzpicture}
    
    \draw (-1.5,-1.5) -- (-1.5,1.5) node [above] {$NS5$};

    \draw (-1.5,0) -- (0.75,0) node [midway,below] {$N_{c} \, D4$};

    \draw (0,-1.5) -- (1.5,1.5) node [above] {$NS5'$};

    \draw (-3.8,0.75) -- (-1.5,0.75) node [midway,below] {$\Nf \, D4$}; 
    
    \node at (-4,0.75){\Large $\otimes$};
   
    \end{tikzpicture}}
    \quad\quad\quad\quad\quad%
    %
    %
    \subfigure[Magnetic Theory]{
    \begin{tikzpicture}
    \draw (1.5,-1.5) -- (1.5,1.5) node [above] {$NS5$};

    \draw (-0.75,0) -- (1.5,0) node [midway,below] {$N_{c} \, \overline{D4}$};

    \draw (-1.5,-1.5) -- (0,1.5) node [above] {$NS5'$};

    \draw (-2.675,0.75) -- (-0.6,0.75) node [midway,below] {$\Nf \, D4$};
    \draw (-0.2,0.75) -- (1.5,0.75) ;
    
    \draw (-2.875,0.9);
    \node at (-2.875,0.75){\LARGE $\otimes$};

	\end{tikzpicture}
    }
    \caption{Including mass for the electric quarks}
    \label{BraneConfiguration2}
	\end{figure}

We can give mass to the quarks in the electric theory by separating the D4 flavour branes from the D4 colour branes in the $(x^{4},x^{5})$ plane. This corresponds to adding the superpotential $W=mQ\bar{Q}$. On the magnetic theory, this corresponds to adding $W=Mm$ to the dual theory action, see \cite{Giveon:1998sr} for a review. The inclusion of the quark masses stabilises the vacuum. The brane configurations with the inclusion of masses is given in figure \ref{BraneConfiguration2} 

\section{Field Theory Content and Anomaly Matching}

The electric theory we consider is SQCD, based on $SU(N_c)$ with $N_f$ flavours, see table \eqref{ChargesElectric2}. The magnetic dual for $N_f<N_c$ can be read from the brane configuration. The gauge theory on the \textit{anti} D4 branes is $SU(N_c-N_f)$. The open strings between the colour and flavour branes give rise to quarks. In addition the open strings that live on the flavours branes lead to a meson in the adjoint of $SU(N_f)$.  The matter content of the magnetic theory is given by the table below.
    \begin{table}[h]
        \centering
        \begin{tabular}{c|cccc}
                  &  $SU(\Nf)_{L}$ & $SU(\Nf)_{R}$ & $U(1)_{B}$ & $U(1)_{R}$  \\ \hline 
            $q$   &     $\Nf$     &    $1$        & $\frac{\Nc}{\Nc-\Nf}$      & $\frac{2\Nf-\Nc}{\Nf}$  \\
            $\bar{q}$ &   $1$   &  $\bar{N}_{f}$   &   -$\frac{\Nc}{\Nc-\Nf}$      & $\frac{2\Nf-\Nc}{\Nf}$  \\
            $M$      &   $\Nf$   & $\bar{N}_{f}$   &   $0$   &   2$\frac{\Nf-\Nc}{\Nf}$
        \end{tabular}
        \caption{Matter content and charges of the Magnetic Theory}
        \label{Magnetic}
    \end{table}

Note that the charge assignation of the magnetic theory for $\Nf<\Nc$ is not the same as in the $\Nf>\Nc$ case, which are listed in \cite{Seiberg:1994pq}. 

While the matter content of the magnetic theory is supersymmetric, the interaction terms break supersymmetry. To be more precise both field theories on the flavour branes and the colour branes are supersymmetric. In additional the colour-flavour branes interaction leads to gauge interactions and Yukawa interactions. The main issue is that in the absence of supersymmetry the squark will acquire a potential. An appealing scenario is that the squark (which is a mode of an open string stretched between the colour anti-branes and the flavour branes) will becomes tachyonic and will condense,  resulting in colour-flavour locking $\langle \phi^i_a \rangle  =v \delta ^i _a$. In the brane picture it means that $N_f$ colour and flavour branes will recombine and move away from the remaining $N_c$ anti-branes. The branes will settle at a position described in figure 3b.

We will find, however, the end point of the RG flow, namely the minimum of the potential where the branes settle.

\subsection{Anomaly Matching}

In order to check the proposed duality beyond the SUSY case, we study the anomalies in both theories. With the charge assignation in tables \ref{ChargesElectric2} and \ref{Magnetic} the anomalies match and are listed below
    \begin{enumerate}
        \item $SU(\Nf)_{L}^{3}$ 
            \begin{itemize}
                \item Electric Theory: $\Nc \Tr(T_{a}T_{b}T_{c})$.
                \item Magnetic Theory: $(\Nc-\Nf)\Tr(T_{a}T_{b}T_{c}) + \Nf \Tr(T_{a}T_{b}T_{c}) = \Nc \Tr(T_{a}T_{b}T_{c})$.
            \end{itemize}
        \item $SU(\Nf)_{L}^{2} \,  U(1)_{R}$
            \begin{itemize}
                \item  Electric Theory: $\Nc\left(\frac{\Nf-\Nc}{\Nf}-1 \right)\Tr(T_{a}T_{b}) = -\frac{\Nc^{2}}{\Nf} \Tr(T_{a}T_{b})$.
                \item Magnetic Theory: $(\Nc-\Nf)\left(\frac{2\Nf-\Nc}{\Nf}-1 \right)\Tr(T_{a}T_{b}) + \Nf\left(2\frac{\Nf-\Nc}{\Nf}-1\right)\Tr(T_{a}T_{b}) = -\frac{\Nc^{2}}{\Nf} \Tr(T_{a}T_{b})$
            \end{itemize}
        \item $SU(\Nf)_{L}^{2} \,  U(1)_{B}$ 
            \begin{itemize}
                \item Electric Theory: $\Nc \Tr(T_{a}T_{b}) $
                \item Magnetic Theory: $(\Nc-\Nf)\frac{\Nc}{\Nc-\Nf}\Tr(T_{a}T_{b}) = \Nc \Tr(T_{a}T_{b})$
            \end{itemize}
        \item $U(1)_{R}$
            \begin{itemize}
                \item Electric Theory: $2\Nf\Nc\left( \frac{\Nf-\Nc}{\Nf}-1\right) + \left( \Nc^{2}-1\right)= -(\Nc^{2}+1)$
                \item Magnetic Theory: $2(\Nc-\Nf)\Nf\left( \frac{2\Nf-\Nc}{\Nf}-1 \right) +\Nf^{2}\left( 2\frac{\Nf-\Nc}{\Nf}-1 \right) + \left( (\Nc-\Nf)^{2}-1\right)= -(\Nc^{2}+1)$
            \end{itemize}
        \item $U(1)^{3}_{R}$
            \begin{itemize}
                \item Electric Theory: $2\Nc\Nf \left( \frac{\Nf-\Nc}{\Nf}-1 \right)^{3} + \left( \Nc^{2}-1\right) =-2\frac{\Nc^{4}}{\Nf^{2}} + \Nc^{2}-1 $
                \item Magnetic Theory: $2(\Nc-\Nf)\Nf\left( \frac{2\Nf-\Nc}{\Nf}-1 \right)^{3} \\ + \Nf^{2}\left(2\frac{\Nf-\Nc}{\Nf}-1 \right)^{3}+ \left( (\Nc-\Nf)^{2}-1\right)= -2\frac{\Nc^{4}}{\Nf^{2}} + \Nc^{2}-1$ 
            \end{itemize}
        \item $U(1)^{2}_{B} U(1)_{R}$
            \begin{itemize}
                \item Electric Theory: $\Nc\Nf 1^{2} \left( \frac{\Nf-\Nc}{\Nf}-1 \right) + \Nc\Nf (-1)^{2} \left( \frac{\Nf-\Nc}{\Nf}-1 \right) = -2\Nc^{2}$
                \item Magnetic Theory: $ (\Nc-\Nf)\Nf \left(\frac{\Nc}{\Nc-\Nf} \right)^{2} \left( \frac{2\Nf-\Nc}{\Nf}-1 \right)  \\
                + (\Nc-\Nf)\Nf \left(-\frac{\Nc}{\Nc-\Nf} \right)^{2} \left( \frac{2\Nf-\Nc}{\Nf}-1 \right) = -2\Nc^{2} $
            \end{itemize}
    \end{enumerate}

\section{M-theory Evidence}

In the M-theory description D4-branes become M5-branes wrapping the 11D circle, which we call $x^{10}$. Defining
    \begin{equation}
        v = x^{4} + i x^{5}, \quad w = x^{8}+ix^{9},\quad s = x^{6} + i x^{10},
    \end{equation}

and using $t = e^{-s/R}$, with $R$ the size of the 11D circle $x^{10}\sim x^{10}+2\pi R$, the description of the electric brane configuration in Type IIA is lifted to a one M5-brane extended in the field theory directions $\mathbb{R}^{3,1}$ and wrapping a two dimensional Riemann surface $\Sigma$ embedded in a 6D space. This Riemann surface corresponds to the Seiberg-Witten Curve for the $\mathcal{N}=1$ theory \cite{Witten:1997ep,Hori:1997ab,Brandhuber:1997iy}. The curve $\Sigma$ is defined by
    \begin{align}
        & t \prod_{i}(v-m_{i}) - \left(\prod_{i}m_{i}\right)^{\frac{\Nf-\Nc}{\Nf}} v^{\Nc} = 0,\label{SWCurve1}\\
        & vw = \left(\prod_{i}m_{i}\right)^{\frac{1}{\Nc}} \Lambda^{\frac{3\Nc-\Nf}{\Nc}},\label{SWCurve2}
    \end{align}
where $m_{i}$ are the quark masses on the electric theory, and $\Lambda$ is the strong interaction scale. Note that we can rewrite \eqref{SWCurve2} using \eqref{vevM} as
    \begin{equation}\label{vwcurve}
        v = \frac{\calD}{w}.
    \end{equation}

The interpretation of \eqref{SWCurve1} goes as follows: as a polynomial in $t$, the degree of the polynomial is related to the number of NS5-branes extended in the $v$ direction. On the other hand, the coefficients of each term of the $t$-polynomial are polynomials in $v$. The zeros of the $v$-polynomial of the $t^{1}$ part give the positions of the D4-branes on the left of the NS5-brane (which in the electric configuration corresponds to the flavour branes) and while the zeroes of the $t^{0}$ part give the number of D4-branes on the right of the NS5-brane. The $N_c$ roots correspond to the number of colours.

From here, in order to obtain the SW curve of the dual magnetic theory we follow the steps outlined in \cite{Schmaltz:1997sq}, where the position of the NS5 branes is interchanged so that flavour branes are placed on the left of the NS5'-brane. The aim is to rewrite the electric curve \eqref{SWCurve1} in order to obtain an equation for the $t-w$ plane of the form
    \begin{equation}
        t P(w) + Q(w) = 0.
    \end{equation}
To this end, we use \eqref{SWCurve2} in \eqref{SWCurve1}, from which we obtain 
    \begin{equation}
        t \prod_{i}\left(\frac{\calD}{w}-m_{i}\right) - \left(\prod_{i}m_{i}\right)^{\frac{\Nf-\Nc}{\Nf}} \left(
            \frac{\calD}{w}\right)^{\Nc} = 0.
    \end{equation}
After a bit of algebra we get (we also shift $t$ to absorb the $(-1)^{\Nf}$ factor)
    \begin{equation}
        t \prod_{i}\left(w-\frac{\calD}{m_{i}}\right) - \left(\prod_{i}\frac{\calD}{m_{i}} \right)^{\frac{\Nc}{\Nf}} w^{\Nf-\Nc} = 0.
    \end{equation}

Note that, from the zeros of the $w$ polynomial of the $t^{1}$ part, we can read the position of the flavour branes on the left of the NS5'-brane, which realises quarks and mesons. Note that the position of the $i$-th brane is $w=\calD/m_{i}$ which is precisely the value of the meson vev in the stabilised vacuum of the electric theory \eqref{vevM}.  In this way, the SW curve in the magnetic language captures information about the IR dynamics of the electric theory.

Finally, the SW curve written in terms of the magnetic variables is
    \begin{equation}
         t \prod_{i}\left(w-\vev{M}_{i}\right) - \left(\prod_{i}\vev{M}_{i} \right)^{\frac{\Nc}{\Nf}} w^{\Nf-\Nc} = 0.
    \end{equation}
The position of the flavour branes on the left of the NS5'-brane now is controlled by $\vev{M}_{i}$.

Note that
\begin{itemize}
    \item If $\Nf>\Nc$ then $\tNc = \Nf-\Nc$, and using these variables the SW curve reads
    \begin{equation}
         t \prod_{i}\left(w-\vev{M}_{i}\right) - \left(\prod_{i}\vev{M}_{i} \right)^{\frac{\Nf-\tNc}{\Nf}} w^{\tNc} = 0 
    \end{equation}
    \item If $\Nf<\Nc$ then $\tNc = \Nc-\Nf$, and the curve reads
    \begin{equation}
         t \prod_{i}\left(w-\vev{M}_{i}\right) - \left(\prod_{i}\vev{M}_{i} \right)^{\frac{\Nf+\tNc}{\Nf}} w^{-\tNc} = 0 
    \end{equation}
\end{itemize}

Due of the factor $w^{-\tNc}$, the equation is not the standard SW curve (in the sense that the polynomials usually have positive powers). 
To understand the consequences of this minus sign, lets analyse the behaviour of the curve near $w\rightarrow 0$. At this point $t\rightarrow \infty$, which corresponds to $x_{6}\rightarrow -\infty$. Furthermore, by \eqref{vwcurve} we have $v\rightarrow \infty$, which corresponds to the location of the NS5-brane. There are also $\Nf$ semi-infinite flavour branes ending on the left of the NS5'-brane, while  the $w^{-\tNc}$ reflects that there are $\Nc-\Nf$ branes stretched from the \textit{left} of the NS5' to the \textit{right} of the NS5. Putting all of this together, the "magnetic" is describing the higgsed phase of the electric theory, where the gauge group is broken to $SU(\Nc)\rightarrow SU(\Nc-\Nf)$. In the discussion section we  explain that the magnetic theory is not a dual of the electric theory in the same sense as for $N_f>N_c$, but rather a manifestation of its IR degrees of freedom.

Similarly, we can rewrite the $v-w$ relation as
    \begin{equation}\label{SWDual2}
        vw = \left(\prod_{i}\vev{M}_{i}\right)^{\frac{1}{\pm\tNc}} \tilde{\Lambda}^{\frac{\pm3\tNc-\Nf}{\tNc}},
    \end{equation}
where the $\pm$ sign corresponds to the $\Nf>\Nc$ and $\Nf<\Nc$ respectively.

As noted in \cite{Schmaltz:1997sq}, even though we just rewrote the Seiberg-Witten curve \eqref{SWCurve1}, \eqref{SWCurve2} and \eqref{SWDual2} also correspond to the Seiberg-Witten curve of the Seiberg dual theory, with gauge group $SU(\tNc)$.

Recall that the SW curve is formulated in an holomorphic language in order to preserve SUSY. For $\Nf<\Nc$, the masses of the quarks allow to have a stable vacuum. This is reflected in the magnetic theory, where in spite of the repulsion between branes and anti-branes there is a finite distance between them, and the SW curve is written in holomorphic variables. On the other hand, from \eqref{vevM} we see that as $m\rightarrow 0$, $\vev{M}_{i}$ admits a runaway behaviour for $\Nf<\Nc$. This is expected since the ADS superpotential does not have a minimum for massless quarks. This result was also obtained in \cite{Hori:1997ab}, where it was shown that if one gives an (infinite) mass to the adjoint scalar in the $\mathcal{N}=2$ $SU(\Nc)$ theory, the position of the flavour branes in the dual curve go to go infinity in the $w$ plane.  This is due to the fact that, in order to write the magnetic curve in holomorphic variables, the curve needs to sit at the vacuum, which corresponds to the branes and anti-branes being infinitely separated.


\section{Discussion: the meaning of the proposed duality}

In order to understand the meaning of the duality let us consider first the electric theory. Let us give a vev to the squarks. For simplicity we consider an identical vev for all squarks 
\begin{equation}
    \langle Q \rangle = \langle \tilde Q \rangle =v
\end{equation}
In the brane language such a vev corresponds to detaching $N_f$ colour branes and reconnecting them to flavour branes. The result is a stack of $N_c-N_f$ colour branes and an addition stack of $N_f$ branes that corresponds to mesons. The distance between the stack of $N_f$ mesons and the the stack of $N_c-N_f$ is $v$. The theory on the colour branes admits a mass gap and a gluino condensate. The gluino condensate gives rise to the ADS superpotential on the stack of $N_f$ 
branes.

Now let us consider the magnetic theory, with $N_c-N_f$ \textit{anti}-branes separated by a distance $v$ from the $N_f$ branes. The magnetic description is almost identical to the electric decsription outlined above: we have a supersymmetric $SU(N_c-N_f)$ gauge theory coupled to $N_f$ mesons. The difference between the electric and magnetic theories is that the electric side is manifestly supersymmetric, whereas the coupling between the branes and antibrans in the magneic side breaks supersymmetry. 

Note that if we swap the fivebranes in the magnetic side the antibranes become branes and we arrive at the electric side. Following the arguments of Elitzur-Giveon-Kutasov, we claim that both the electric and magnetic theories flow to the same IR theory. In other words: the interactions that violate the equivalence between the electric and magnetic theories in the UV become irrelevant in the IR. 

In conclusion we propose that the magnetic theory describes in the IR the ADS superpotential. The advantage of the magnetic description is that mesons are elementary fields and the IR degrees of freedom are manifest.

The proposed duality for $N_f<N_c$ is weaker than the duality for $N_f>N_c$\footnote{We would like to thank Nathan Seiberg for a useful discussion on this topic.}. The original duality of Seiberg reveals the emergence of new degrees of freedom in the IR, in particular massless gauge bosons. The current duality is a tool to reveal results that were already obtained in the electric language. 

Nevertheless the duality is useful: as we already saw, it leads to interesting new results upon reduction to 3d and it may lead to a better understanding of QCD dyanmics, especially since in the magnetic dual the meson field is elementary.

A further interesting direction of research is to study how the 3d duality emerges from 4d upon dimensional reduction. For the $N_f>N_c$ case the reduction was carried out in ref.\cite{Aharony:2013dha}.\\ \\

{\bf Acknowledgments} We thank Nick Dorey, Carlos Hoyos, Zohar Komargodski, Nathan Seiberg and Shigeki Sugimoto for very useful discussions. RS acknowledges support from STFC grant ST/W507878/1. We are supported by  STFC grant ST/T000813/1.

{\footnotesize {\bf Open Access Statement} - For the purpose of open access, the authors have applied a Creative Commons Attribution (CC BY) licence to any Author Accepted Manuscript version arising. 

{\bf Data access statement}: no new data were generated for this work.}

\bibliographystyle{JHEP}

\bibliography{ref}


\end{document}